# Building Digital Twins of Different Human Organs for Personalized Healthcare

**Yilin Lyu[1,#], Zhen Li[1,#], Vu Tran[1], Xuan Yang[1], Hao Li[1], Meng Wang[2,3], Ching-Yu Cheng[2,3,4,5], Mamatha Bhat[6,7,8,9], Viktor Jirsa[10], Roger Foo[11,12,13], Chwee Teck Lim[1,14,15], Lei Li[1,*]**

[1] Department of Biomedical Engineering, National University of Singapore, Singapore, Singapore

[2] Centre for Innovation and Precision Eye Health, Yong Loo Lin School of Medicine, National University of Singapore, Singapore, Singapore

[3] Department of Ophthalmology, Yong Loo Lin School of Medicine, National University of Singapore, Singapore, Singapore

[4] Singapore Eye Research Institute, Singapore National Eye Centre, Singapore, Singapore

[5] Ophthalmology & Visual Sciences Academic Clinical Program (EYE ACP), Duke-NUS Medical School, Singapore, Singapore

[6] Ajmera Transplant Centre, Toronto General Hospital, University Health Network, Toronto, ON, Canada

[7] Toronto General Hospital Research Institute, University Health Network, Toronto, ON, Canada

[8] Vector Institute, Toronto, ON, Canada

[9] Department of Medicine, University of Toronto, Toronto, ON, Canada

[10] Aix Marseille University, INSERM, INS, Inst Neurosci System, Marseille, France

[11] Cardiovascular-Metabolic Disease Translational Research Programme, Yong Loo Lin School of Medicine, National University of Singapore, Singapore, Singapore

[12] Department of Cardiology, National University Heart Centre, Singapore, Singapore, Singapore

[13] Cardiovascular Research Institute (CVRI), Yong Loo Lin School of Medicine, National University of Singapore, Singapore, Singapore

[14] Institute for Health Innovation and Technology (iHealthtech), National University of Singapore, Singapore, Singapore

[15] Mechanobiology Institute, National University of Singapore, Singapore, Singapore

[#] These authors contributed equally

[*] Corresponding author: lei.li@nus.edu.sg

## ABSTRACT

**Digital twins are virtual replicas of physical entities and are poised to transform personalized medicine through the real-time simulation and prediction of human physiology. Translating this paradigm from engineering to biomedicine requires overcoming profound challenges, including anatomical variability, multi-scale biological processes, and the integration of multi-physics phenomena. This survey systematically reviews methodologies for building digital twins of human organs, structured around a pipeline decoupled into anatomical twinning (capturing patient-specific geometry and structure) and functional twinning (simulating multi-scale physiology from cellular to organ-level function). We categorize approaches both by organ-specific properties and by technical paradigm, with particular emphasis on multi-scale and multi-physics integration. A key focus is the role of artificial intelligence (AI), especially physics-informed AI, in enhancing model fidelity, scalability, and personalization. Furthermore, we discuss the critical challenges of clinical validation and translational pathways. This study not only charts a roadmap for overcoming current bottlenecks in single-organ twins but also outlines the promising, albeit ambitious, future of interconnected multi-organ digital twins for whole-body precision healthcare.**

## 1. Introduction

Digital twins are now widely seen as a transformative force in healthcare, promising to revolutionize personalized medicine by enabling real-time monitoring, predictive diagnostics, and optimized treatment strategies[1,2]. However, the transition from a promising concept to clinically viable tools remain a formidable challenge. This challenge requires moving beyond two dominant paradigms: conventional computational modeling, which often personalizes anatomy but relies on static, literature-derived functional parameters[3], and general-purpose AI, such as foundation models, which have garnered significant attention for broad pattern recognition but often prioritize generality over the deep personalization and interpretability essential for clinical trust[4-7]. The digital twin paradigm proposes a distinct, complementary solution. It is a technology fundamentally oriented toward building a patient-specific, data-assimilating, and dynamically evolving virtual replica. This is achieved by continually integrating individual clinical data with well-established mathematical and physical models, such as the Navier-Stokes equations for hemodynamics[8] or the Monodomain/Bidomain models for cardiac electrophysiology[9]. By grounding predictions in known biophysical principles, the digital twin framework aims to deliver inherently personalized, interpretable, and actionable insights, directly addressing the limitations of purely data-driven or static modeling approaches[10,11].

Constructing such high-fidelity, organ-specific digital twins confronts the immense complexity of human physiology. Unlike their industrial counterparts, which operate in controlled environments, biological digital twins must account for intrinsic patient variability, multi-scale interactions (from molecules to whole-organ function), and the integration of sparse, noisy, or incomplete clinical data[15-18]. This complexity reveals limitations in both foundational modeling approaches. While physics-based models provide essential mechanistic insight and interpretability, they often struggle with computational cost and incomplete biological knowledge. Purely data-driven AI models, conversely, may fail under data-scarce conditions and lack the interpretability required for clinical adoption[19]. To bridge this divide, hybrid approaches that synergistically combine mechanistic modeling with machine learning are emerging as the most promising technical path forward[20]. In such frameworks, well-understood physical laws are encoded into the model architecture (e.g., via physics equations), while data-driven components are tasked with learning the unknown or residual physics that are difficult to model explicitly. However, the precise methodologies for constructing, validating, and deploying these integrated, clinically reliable systems are still nascent. A critical gap persists between the high-level vision and practical implementation, underscoring the urgent need for a systematic synthesis of the technical frameworks required to realize functional digital twins in healthcare.

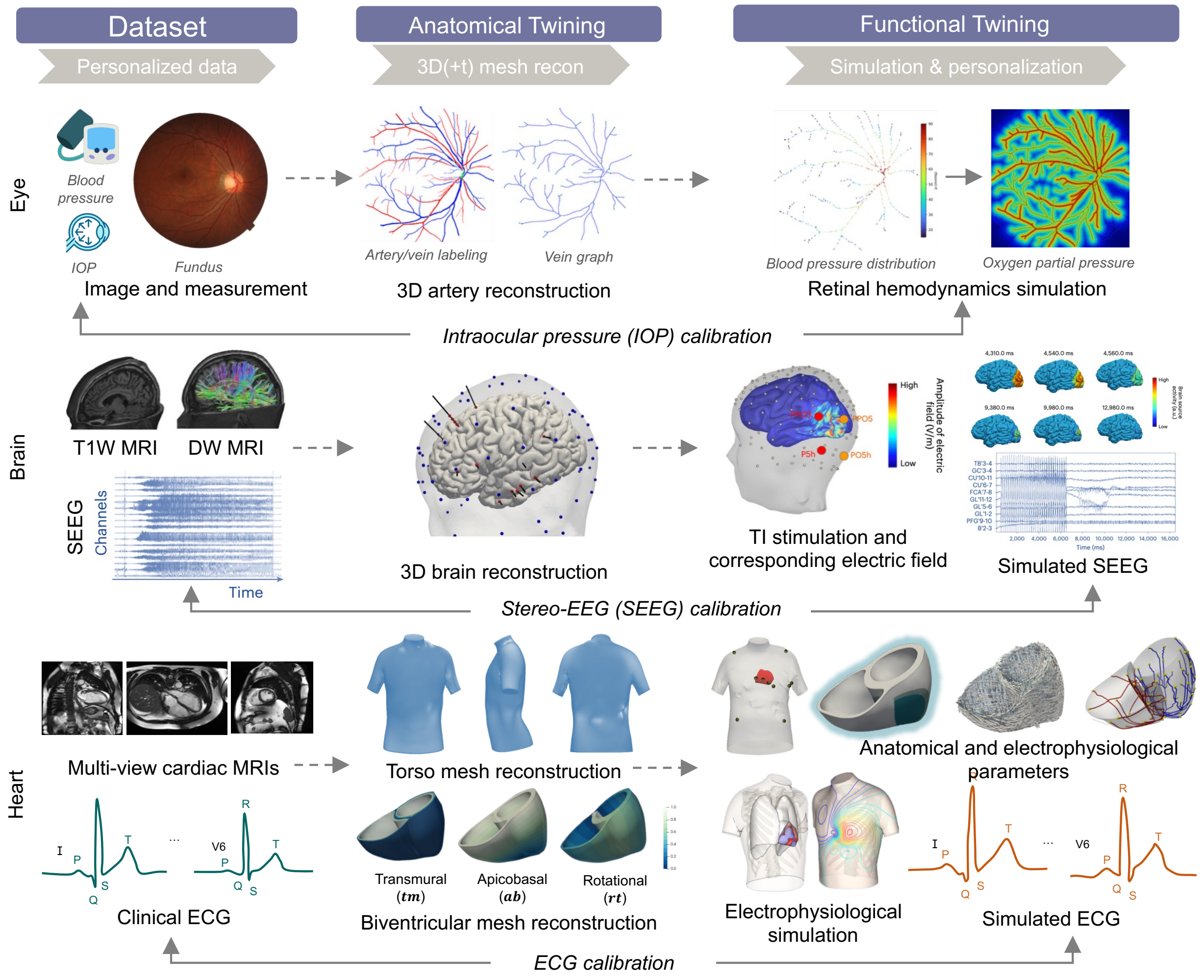

**Fig. 1**: Pipeline for developing organ-specific digital twins, including anatomical and functional twinings based on patient specific dataset. Here, we use the digital twin of the eyes[12], brain[13], and heart[14] as examples. Images adapted with permission from ref. 13-14 under a Creative Commons license CC BY 4.0.

This survey systematically reviews state-of-the-art computational frameworks for constructing digital twins of different human organs. In general, the digital twin workflows usually involve two stages, namely anatomical and functional twinnings, as presented in Fig. 1. The anatomical twinning stage involves the segmentation of medical images, reconstruction of the three-dimensional (3D) or 3D+t geometry of the organ, and the localization of relevant pathologies. At the functional twinning stage, the mathematical models are used to simulate physiological function of the organ. More importantly, the parameters of the mathematical models need to be optimized via inverse inference using patient-specific data for personalization. In this survey, we examine the methodologies for both anatomical and functional modeling and the key challenges in terms of methodologies. By comparing methodologies across different organs, we provide a roadmap for advancing digital twins from conceptual promise to clinical reality.

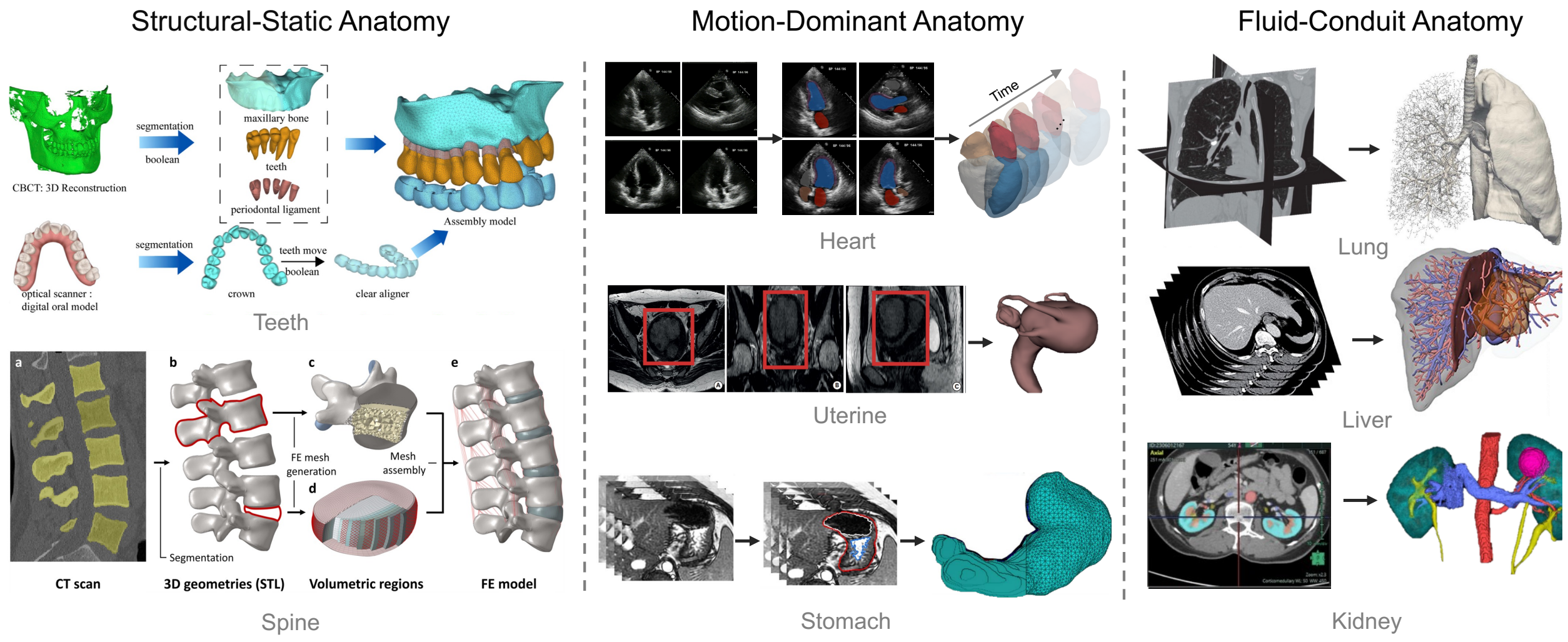

**Fig. 2**: Overview of digital twin anatomical modeling of different anatomy types. The pipeline is shaped by the dominant biomechanical behavior of organs: (1) Structural-Static Anatomy (e.g., teeth[31], spine[32]) demands high geometric accuracy for structural planning. (2) Motion-Dominant Anatomy (e.g., heart, uterus[33], stomach[34]) requires 4D spatiotemporal reconstruction to capture deformation. (3) Fluid-Conduit Anatomy (e.g., lung[35], liver[36], kidney[37]) focuses on extracting topologically correct, connected tree structures for flow simulation. Images adapted with permission from refs. 31-33, 35-37 (CC BY 4.0) and ref. 34 (CC BY-NC 4.0).

## 2 Organ-Specific Anatomical Twining

The development of clinically viable digital twins necessitates a foundational anatomical model that serves not merely as a visual representation, but as a structured, simulation-ready computational domain. This model, derived from patient-specific imaging data, forms the essential geometric and topological scaffold upon which biophysical laws are solved. The fidelity and utility of any subsequent functional modeling are therefore fundamentally predicated on the anatomical model. This pipeline for anatomical modeling traditionally involves a two-stage process: 1) the semantic segmentation of anatomical structures from 2D/3D(+t) medical images, and 2) the conversion of these discrete labels into continuous, high-quality geometric meshes suitable for numerical simulation.

The segmentation stage has been revolutionized by deep learning. Following early reliance on manual delineation and classical methods (e.g., graph-cut[21], multi-atlas segmentation[22]), convolutional neural networks, particularly U-Net architectures[23] and their optimized variants especially nnU-Net[24], have become the dominant paradigm. More recently, transformer-based models (e.g., TransUNet[25]) and medical-adapted foundational segmentation models (e.g., MedSAM[26]) are pushing the boundaries of accuracy and generalization across diverse imaging modalities and anatomical structures.

The subsequent mesh generation stage, however, remains a significant computational bottleneck. Converting segmented labels into simulation-ready meshes, free of topological errors, with appropriate element quality, and preserving physiological interfaces, is non-trivial. Traditional isosurfacing (e.g., Marching Cubes) followed by volumetric meshing often propagates segmentation artifacts and requires extensive manual correction[27]. Emerging end-to-end and learning-based approaches aim to bypass these limitations[28]. Techniques leveraging differentiable rendering[29], graph convolutional networks[30], and direct image-to-mesh architectures[28] represent a paradigm shift toward generating physics-compatible geometries directly from imaging data. Critically, the design priorities for this anatomical scaffold are not uniform across all organs. They are intrinsically dictated by the dominant biophysical processes the digital twin aims to simulate. We posit that organ-specific anatomical modeling strategies can be effectively classified into three computational paradigms: structural-static anatomy, motion-dominant anatomy, and vascular/parenchymal network anatomy (Fig. 2). Here, the geometry of structural-static organs is largely invariant, and the primary modeling goal is precise structural representation for intervention planning or as a stable mechanical scaffold. The primary function of motion-dominant organs involves cyclic or aperiodic deformation, necessitating spatiotemporal (4D) reconstruction to capture dynamic geometry. For organs whose function is defined by the transport of fluids through a hierarchical branching tree (blood vessels, ducts, airways), topological connectivity is crucial. This classification is not merely a descriptive convenience; it is a prescriptive framework that guides the prioritization of modeling features, the selection of computational algorithms, and the very criteria for validating the anatomical pipeline, ensuring that the model is fit-for-purpose from its inception.

## 2.1 Static and Semi-Static Structural Anatomy

This paradigm encompasses organs and tissues whose geometry, for the purposes of the digital twin, can be treated as fixed or changing only due to externally imposed interventions (e.g., surgery, growth). The computational objective is to achieve a high degree of geometric accuracy to serve as a template for prosthetic design, surgical planning, or as a stable reference frame for functional simulations.

Modeling the musculoskeletal system and skin presents a unique computational contrast between rigid and deformable tissues. Bone modeling, primarily from computed tomography (CT), is a mature domain where threshold-based segmentation yields highly accurate surfaces for 3D printing and surgical guide fabrication[38]. The challenge shifts to muscle and skin, which are deformable and require preservation of volume and biomechanical consistency. Early physics-based methods utilized implicit surfaces[39] and volume-preserving constraints (e.g., volumetric Laplacian formulations[40]). Contemporary approaches integrate data-driven techniques with physical priors. For instance, Li et al.[41] combined Neo-Hookean hyperelasticity constraints with iterative shape registration for subject-specific muscle deformation. Multiscale modeling, such as using deep generative networks to synthesize trabecular bone microstructure from micro-CT data[42], exemplifies the push towards hierarchical anatomical fidelity. Skin modeling adds the complexity of layered structure and optical properties. While finite element models of epidermal/dermal layers have been used to study wound healing[43], recent advances leverage 3D photogrammetry and optical coherence tomography (OCT) for high-fidelity surface and subdermal reconstruction[44], with deep learning enabling automatic segmentation from multi-view images[45].

Although brain function is dynamic, its gross anatomical architecture for many large-scale simulation frameworks (e.g., The Virtual Brain[46]) is often treated as a semi-static conductive domain. Traditional neuroimaging pipelines rely on atlas-based tools like FreeSurfer[47] and FSL[48] to segment gray matter, white matter, and subcortical structures from MRI, followed by surface and volumetric meshing. The propagation of segmentation errors to meshes remains a key challenge[49]. This has motivated the development of accelerated deep learning replicants of traditional pipelines (e.g., FastSurfer[50]) and, more importantly, end-to-end methods that directly infer topologically correct surfaces. Architectures like Vox2Cortex[51], CortexODE[52], and CorticalFlow[53] use neural ODEs and flow-based models to generate cortical surfaces with inherent correspondences, significantly reducing error accumulation and processing time.

The modeling of craniofacial and aural structures, such as teeth, jaw, and ear, forms a subdomain defined by integrating extreme geometric precision with aesthetic and functional demands[54-56]. This unifies distinct sites into a coherent challenge: dental twins require multi-modal fusion of structural and color data for prosthetics; outer ear models achieve sub-millimeter accuracy via scanning and template morphing for hearing aids; while middle ear modeling confronts inaccessible micro-anatomy, often relying on HRCT and manual refinement[57,58]. Across these applications, the computational objective extends beyond anatomical representation to a level of precision that directly dictates clinical utility and patient outcome.

## 2.2 Motion-Dominant Anatomy

This paradigm addresses organs whose primary physiological function is intrinsically linked to large-scale, cyclic deformation. The anatomical model must therefore be a 4D reconstruction that accurately captures the spatiotemporal trajectory of the organ, imposing strict requirements on dynamic imaging, temporal registration, and deforming mesh generation.

Cardiac anatomical twinning inherently demands 4D anatomical modeling to faithfully capture the cardiac cyclic deformation, valve motion, and dynamic blood flow throughout the cardiac cycle via cine MRI, 3D(+t)-CT, or 2D/3D(+t) ultrasound[59,60]. A central computational challenge is reconstructing high-fidelity shapes from imaging data with limited through-plane resolution within cine MRIs or 2D(+t) echo[61,62], often addressed through multi-view learning[63] or the use of sparse point clouds from sequential slices[64]. Recent frameworks like Mesh4D[65] employ transformer-based variational autoencoders to jointly learn shape and motion from multi-view image sequences, producing temporally coherent 4D meshes essential for simulating electrophysiology, biomechanics, and hemodynamics.

Uterine anatomy is highly variable and changes with physiological state (e.g., menstrual cycle, pregnancy)[66]. Reconstruction from 3D transvaginal ultrasound[67] or MRI[68] provides a static snapshot, but functional twins for electrophysiology (e.g., modeling uterine contractions) or biomechanics (e.g., simulating labor) require the incorporation of tissue anisotropy. This involves stratifying the myometrium into layers and assigning characteristic fiber orientations (inner circumferential, outer longitudinal) based on anatomical rules[69,70] or microstructural imaging[71]. Furthermore, comprehensive models often extend to include interactions with pelvic floor muscles and ligaments to study support dynamics and parturition[72,73].

As a central digestive organ, the stomach undergoes complex peristaltic and accommodative deformations. Anatomical modeling must therefore capture its highly variable morphology, influenced by food content and motility phase[74]. Strategies include fusing volumetric CT data with high-resolution endoscopic surface reconstructions obtained via structure-from-motion[75]. The resulting geometries are used in computational fluid dynamics (CFD) models to study gastric mixing[76] and are coupled with biomechanical representations (e.g., fiber-reinforced visco-hyperelastic laws[77]) to simulate wall stress and accommodation.

While the airway network of the lung falls under fluid-conduit anatomy, its parenchymal tissue is motion-dominant, deforming with respiration. Reconstruction must be explicitly correct for respiratory and cardiac motion artifacts in CT/MRI[78]. Techniques like attention-augmented Long Short-Term Memory (LSTM) networks[79] and motion-compensated iterative reconstruction[80] aim to produce consistent, phase-resolved geometric models. Similarly, the eye exhibits dynamic accommodation. While global ocular shape can be reconstructed from fundus photos[81] or MRI[82], modeling accommodation requires parameterized representations of the lens and ciliary body. Consequently, computer-aided design (CAD) based parametric models are often utilized to generate finite element meshes for coupled optical-biomechanical simulation[83].

### 2.3 Fluid-Conduit Anatomy

Organs in this category are defined by their intricate, hierarchical fluid transport systems (arteries, veins, ducts, airways). The main computational objective is the extraction of a topologically correct, connected tree structure that faithfully represents branching patterns down to physiologically relevant scales.

Digital twins of the liver and kidney require dual reconstruction: the parenchymal volume and the complex vascular trees (hepatic artery/portal vein/veins; renal artery/vein/pelvicalyceal system). Contrast-enhanced CT or MRI allows segmentation of major vessels, but the complete vascular network is typically incomplete. Computational workflows thus involve centerline extraction, skeletonization, and algorithmic reconstruction using physiological principles (e.g., Murray's Law) to generate missing branches[84]. For the kidney, vascular reconstruction is particularly critical for surgical planning (e.g., partial nephrectomy)[85]. Significant research focuses on improving the segmentation of thin, low-contrast vessels using topology-aware loss functions[86], boundary-focused metrics[87], and cross-domain validation frameworks[88]. The resulting outputs are 1D graph models for fast hemodynamics or 3D tubular meshes for high-fidelity CFD.

The lung presents one of the most complex branching networks in the human body. Traditional reconstruction methods frequently fail to capture the full, fractal nature of the airway tree, leading to premature truncation[89]. Deep learning approaches have shown promise in preserving distal bronchiole morphology under noisy conditions[90]. The parallel challenge of reconstructing the pulmonary vasculature is equally demanding, with similar requirements for topology preservation to accurately model gas exchange and perfusion.

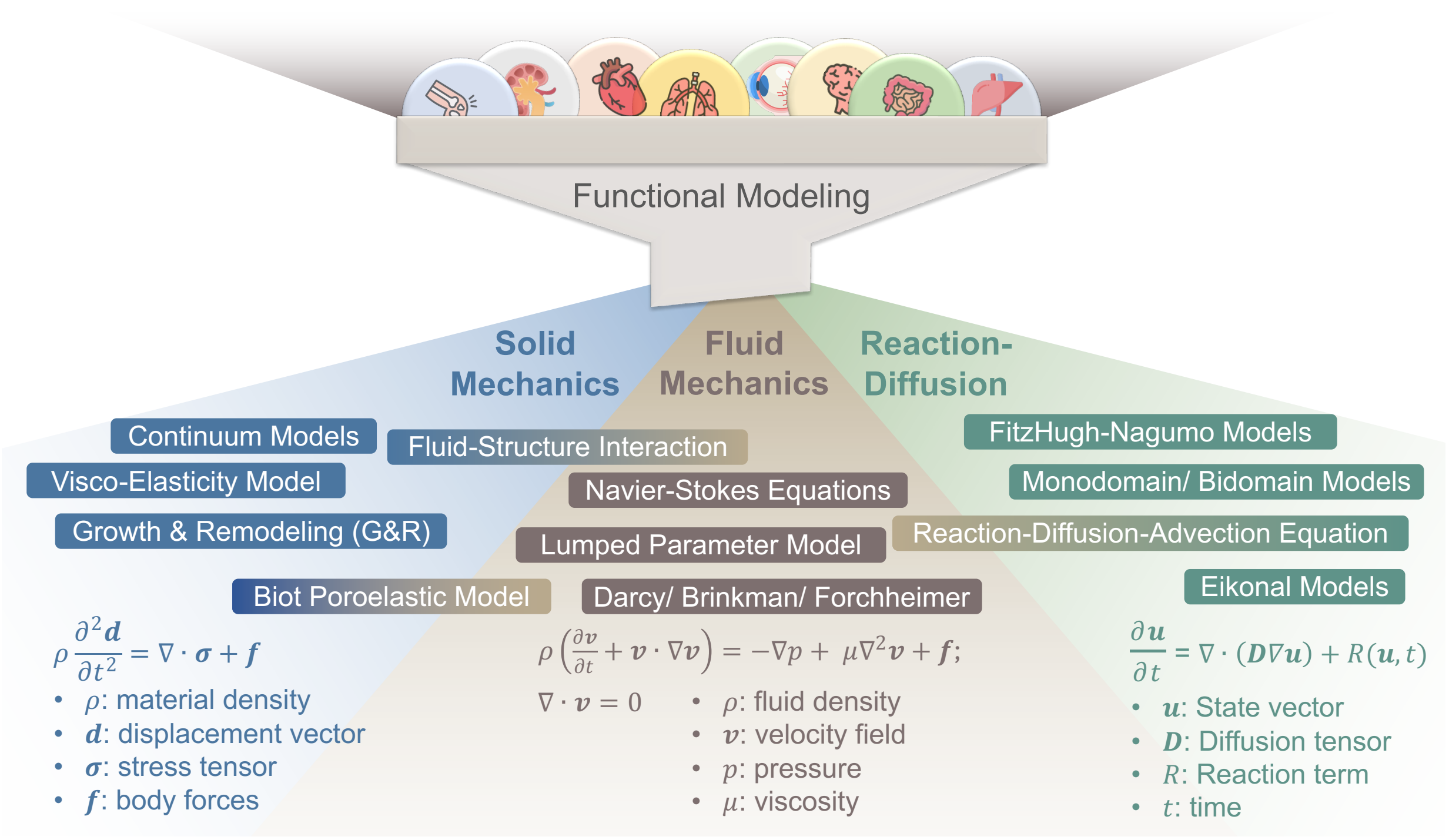

**Fig. 3**: Three major physics-based paradigms for digital twin functional modeling, each defined by a typical partial differential equation. Each paradigm underpins several established computational models, and few advanced models integrate multiple paradigms for greater fidelity.

Beyond the parenchymal geometry, cerebrovascular networks, segmented from magnetic resonance angiography (MRA) or CT angiography (CTA), are essential for hemodynamic models of stroke, aneurysm rupture, and neurovascular coupling. Furthermore, the structural connectome, representing the network of white matter fiber tracts, is a cornerstone of large-scale brain network simulations. It is reconstructed non-invasively using diffusion MRI tractography pipelines (e.g., MRtrix3[91], TractSeg[92]), where the accuracy of the underlying anatomical model directly influences the estimated connectivity weights and subsequent simulation outcomes[93].

## 3 Organ-Specific Functional Twining

The transition from a static anatomical scaffold to a dynamic digital twin requires integrating mathematical models that simulate physiological function, bridging the gap between structural data and physiological behavior in computational medicine[94]. The field has matured through a strategic focus on mastering the dominant physical processes that govern the primary function of each organ, leading to the emergence of three core computational paradigms (Fig. 3). The first is the modeling of electrophysiology and signal propagation (Reaction-Diffusion), which captures the dynamics of excitable tissues and biochemical networks[95]. The second encompasses fluid and hemodynamics, addressing the convective and diffusive movement of biological fluids[96]. The third is structural biomechanics, which simulates the mechanical response of tissues to force[97]. While many state-of-the-art studies excel within a single paradigm due to computational and methodological constraints, the ultimate frontier lies in their integration to capture emergent, multi-physics physiology[98].

### 3.1 Electrophysiology and Signal Propagation Simulation

A fundamental class of organ function arises from the generation and propagation of electrochemical signals, a domain governed by the mathematics of reaction-diffusion systems and network theory. The brain stands as the most complex embodiment of this paradigm[46]. At one extreme, detailed models of spiking neural networks strive for biophysical realism at the cellular level[100]. At the other extreme, mean-field models, such as those implemented in The Virtual Brain platform, abstract the collective activity of neuronal populations into a limited number of state variables[93]. A critical advancement is the development of Bayesian inversion techniques, which enable the estimation of hidden neurophysiological parameters from noninvasive electrophysiological data, demonstrating significant clinical translatability in epilepsy[101].

The heart represents the most mature application of electrophysiological modeling, with a clear pathway to clinical impact in arrhythmia management[95,102]. The modeling pipeline is rigorously multi-scale, beginning with high-fidelity mathematical representations of cardiac myocyte ion channels, such as the Luo-Rudy models[103]. These cellular dynamics are then incorporated into the monodomain or bidomain equations, which are continuum approximations for electrical wave propagation in cardiac tissue[10]. The creation of a patient-specific cardiac electrophysiology twin hinges on solving the ill-posed inverse problem of determining the spatial distribution of tissue properties from body-surface electrocardiogram (ECG) or invasive electrode mapping data[104].

Beyond the central nervous and cardiovascular systems, electrophysiological principles underpin the function of other rhythmic organs. The uterus, for instance, exhibits unique, state-dependent dynamics modulated by hormonal cycles and mechanical stretch[105]. Computational models of uterine electrophysiology capture the specific ionic currents of uterine smooth muscle[106], with upscaling often addressed through reaction-diffusion frameworks or cellular automata approaches[70,107]. Similarly, the motility of the stomach is coordinated by slow-wave activity generated by interstitial cells of Cajal, simulated using reaction-diffusion models to understand dysrhythmias[108]. Finally, the paradigm extends to endocrine organs like the pancreas and thyroid, where the core functional models are systems of ordinary differential equations rep-resenting hormonal feedback loops[109,110].

### 3.2 Fluid and Hemodynamics Simulation

A second major paradigm addresses the critical physiological role of fluid, encompassing the flow of blood, air, and other bodily fluids. The core physics are derived from the Navier-Stokes equations, with a hierarchy of models employed to balance fidelity against computational cost across vast spatial scales[96]. In the cardiovascular system, 3D CFD is used to resolve complex, unsteady blood flow in patient-specific geometries, providing detailed maps of hemodynamic wall shear stress[111]. For efficient simulation of system-wide hemodynamics, zero-dimensional (0D) lumped-parameter models normally represent the global impedance of the vascular network[112]. The lungs present a unique challenge where airflow must be matched with blood flow for efficient gas exchange (Fig. 4c). CFD models of the airway tree simulate ventilation[113], while multi-scale vascular networks model perfusion[114]. The critical coupling of ventilation-perfusion (V/Q) ratios is fundamental to understanding gas exchange efficiency in diseases like COPD and ARDS[115]. The AirPROM project represents a comprehensive effort to integrate these components into a unified computational platform[116].

In the abdominal organs, fluid dynamics integrates with metabolic function. The filtration and metabolic processing of the liver are modeled using porous media approaches for sinusoidal blood flow coupled with reaction-diffusion systems for hepatocyte metabolism[117,118]. Similarly, the kidney employs multi-scale hemodynamic models that incorporate autoregulatory mechanisms like tubuloglomerular feedback[119], with advection-diffusion-reaction equations simulating solute transport in the nephron[120].

Specialized fluid dynamics applications extend to other organs. Ocular models simulate aqueous humor dynamics to understand glaucoma pathophysiology[83]. Cochlear models employ CFD to simulate sound-induced fluid motion in the inner ear[121]. Prostate models use urethral CFD to assess urinary flow dynamics in benign prostatic hyperplasia[122]. The exocrine pancreas models ductal fluid and bicarbonate secretion using transport equations[123].

### 3.3 Solid Biomechanics and Tissue Mechanics Simulation

The third paradigm focuses on the mechanical integrity, deformation, and force-generation of organs and tissues. The musculoskeletal system represents a primary application domain, where bone mechanics are simulated using finite element method (FEM) derived from CT scans to assess fracture risk and design orthopedic implants[97]. This structural analysis is complemented by kinematic simulations that use inverse dynamics to calculate muscle and joint forces during movement, personalizable with motion capture and electromyography data[124]. Dental biomechanics presents unique challenges in modeling hierarchical composite structures. FEM models of teeth must account for the anisotropy of enamel and the non-linear, viscoelastic behavior of the periodontal ligament[125]. These models are further enhanced through mechanobiological algorithms that simulate bone remodeling in response to orthodontic forces[126].

In soft tissue biomechanics, ocular models employ FEM to compute stress-strain distributions in the cornea and sclera, personalizing predictions for glaucoma management and refractive surgery outcomes[83]. The middle ear uses multi-body dynamics and FEM to simulate sound transmission through the ossicular chain[127]. Skin models simulate mechanical behavior for wound healing and surgical planning, incorporating complex hyperelastic and viscoelastic material properties[43].

### 3.4 Multi-Physics Simulation

While significant advances have been made within individual modeling paradigms, the most profound challenge and opportunity lie in their integration. Cardiovascular system provides the most advanced example, where coupled electro-mechanical-fluid models represent the gold standard for comprehensive cardiac digital twinning[98] (Fig. 4b). This integration involves linking electrophysiology models that generate activation sequences with active mechanics models that simulate contraction, which then drive hemodynamic models of blood flow in a dynamically deforming geometry. Similar coupling challenges exist across other organ systems: integrating lung tissue mechanics with airflow and blood flow[128], linking uterine electrophysiology with intrauterine pressure[107], and connecting liver hemodynamics with lobular-scale metabolism[129]. The concept of the Virtual Physiological Human (VPH)[130] emphasized the integration from molecular, cellular, tissue, organ to whole-body levels. It originated in 2005 and aims to establish a comprehensive physiological model of the whole body through multi-scale and multi-physics approaches. Although there was no digital twin at this time, its functions and objectives were similar, embodying the fundamental prototype of digital twin.

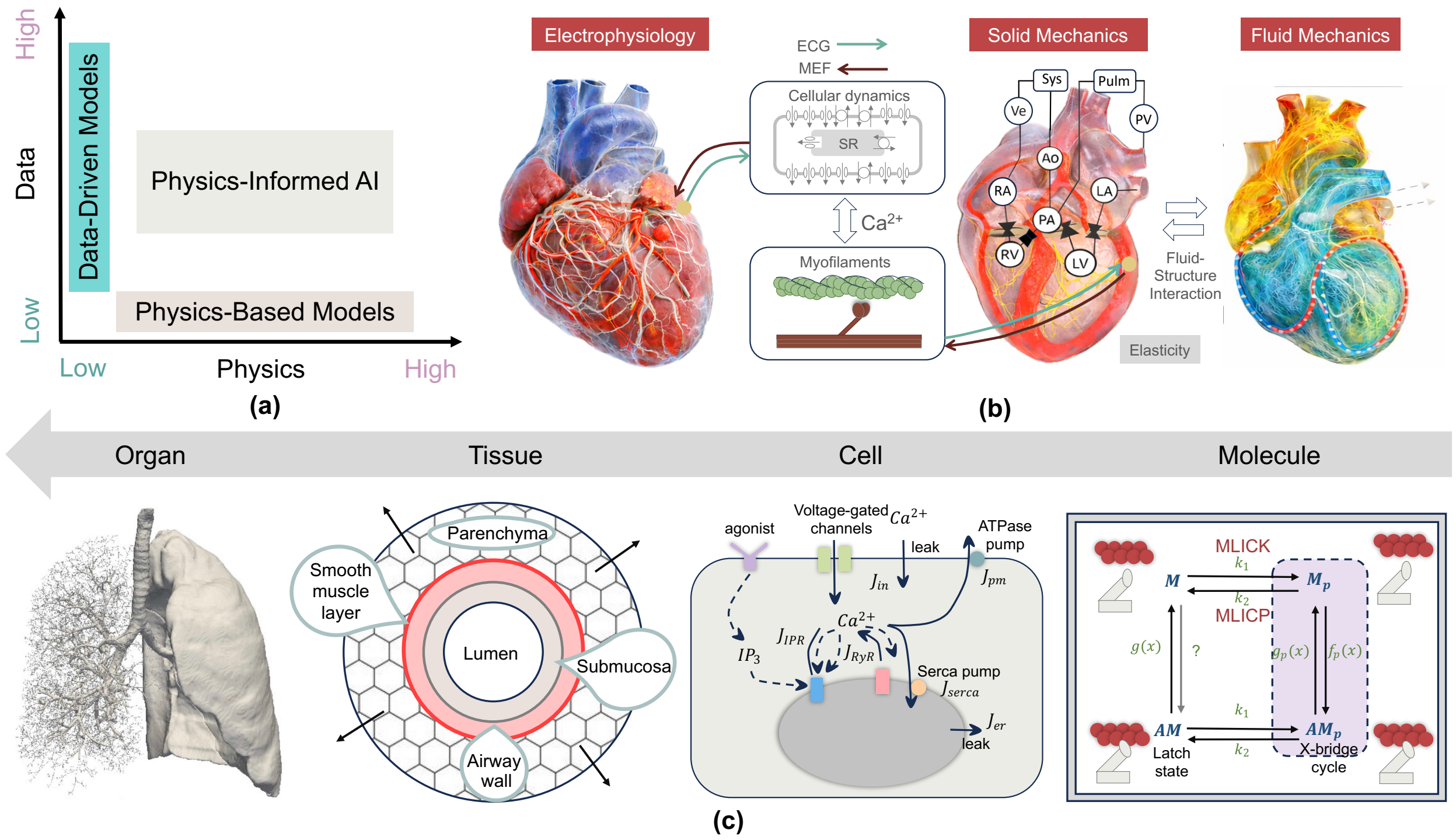

**Fig. 4**: Functional twining approaches: (a) Physics-informed AI model unifying data-driven and physics-based approaches; (b) Multi-physics simulation, exemplified by cardiac electro-fluid-mechanics; (c) Multi-scale modeling, illustrated with a lung digital twin referring to Lauzon et al.[99] under a Creative Commons license CC BY-NC 4.0.

The path forward requires addressing fundamental computational bottlenecks through advanced numerical methods and model reduction techniques. Machine learning approaches are emerging as powerful tools for creating efficient surrogates of complex physics-based models and for solving challenging inverse problems[131]. Furthermore, the development of robust validation frameworks and uncertainty quantification methods will be crucial for building clinical confidence in digital twin predictions. As these challenges are addressed, the field will progressively evolve from single-organ, single-physics models toward comprehensive, multi-scale, multi-physics representations of human physiology, ultimately realizing the vision of truly predictive digital twins in clinical practice.

## 4 Discussion

### 4.1 Physics-Informed AI Powered Digital Twins

The integration of AI with mechanistic modeling is catalyzing a paradigm shift in digital twin development, moving beyond purely data-driven or purely physics-based approaches toward hybrid frameworks that enforce physical and biological constraints (Fig. 4a). Physics-Informed Neural Networks (PINNs) and their variants exemplify this trend by embedding governing equations (e.g., Navier-Stokes, monodomain, or continuum mechanics equations) directly into the loss function of a neural network[134,135]. This allows for the solution of forward and inverse problems even with sparse, noisy clinical data, effectively learning a solution that respects the underlying biophysical laws. For organ digital twins, this approach is particularly powerful for calibrating patient-specific model parameters from heterogeneous observations, such as estimating regional myocardial stiffness from cardiac MRI strain data or inferring blood flow parameters from sparse 4D flow MRI[136,137]. Moreover, operator learning frameworks, such as Fourier neural operators (FNOs)[138] and deep operator networks (DeepONets)[139], can learn mappings between infinite-dimensional function spaces, enabling rapid emulation of complex physiological simulations (e.g., predicting hemodynamics across a family of vascular geometries) at a fraction of the computational cost of traditional solvers. These AI-powered surrogates are essential for enabling real-time digital twin applications, such as surgical planning or closed-loop control in critical care.

However, the clinical translation of physics-informed AI digital twins faces significant computational and methodological bottlenecks. A primary limitation is geometry generalization: standard PINN implementations are typically trained on a single, fixed anatomical mesh. Retraining a new network for the unique geometry of each patient is computationally prohibitive, negating the efficiency gains sought through AI. This has spurred research into geometry-agnostic operator learning. For instance, Yin et al.[140] introduced the diffeomorphic mapping operator learning framework, which learns solution operators for PDEs across diverse geometries. Critically, their work demonstrated scalable application to personalized cardiac electrophysiology, predicting electrical propagation on thousands of cardiac digital twins without retraining, a key capability for clinical deployment. Similarly, graph neural networks operating on un-structured meshes[141] and implicit neural representations[142] are being explored to handle complex, patient-specific topologies. Beyond generalization, the computational cost of solving coupled, multi-physics problems (e.g., cardiac electromechanics or hepatic perfusion) in 3D remains a challenge, often leading to unstable training and high memory demands. Future progress hinges on developing robust, multi-scale training schemes, hybrid methods that strategically blend AI emulators with traditional solvers at critical domains and leveraging multi-fidelity data. Overcoming these geometry-transfer and scalability barriers is essential to move from proof-of-concept demonstrations to robust, patient-specific digital twins that operate within clinically relevant timeframes.

### 4.2 Towards Integrated Multi-Organ Digital Twins

Developing multi-organ digital twins is crucial for capturing the complex inter-organ interactions that underlie systemic diseases[143]. Stroke represents a prime example, where a comprehensive digital twin should encompass the entire heart-brain axis to enable mechanistic insights and personalized decision support[144]. Such a model would integrate blood flow and thrombosis dynamics across the left heart, carotid arteries, and cerebral circulation. By further incorporating brain perfusion, metabolism, and cerebrospinal fluid physiology, it can predict penumbra evolution, optimize intervention timing, and assess risks of reperfusion injury and cerebral edema, enabling personalized neuroprotection and recovery strategies. Similarly, cardiac arrest often leads to multi-organ injury, most notably affecting the brain and kidneys[145,146]. Therefore, developing a digital twin to investigate cardiac arrest and predict its systemic consequences must involve a multi-organ framework that integrates cardiac, cerebral, and renal models to capture the complex cross-organ dynamics driving post-arrest outcomes.

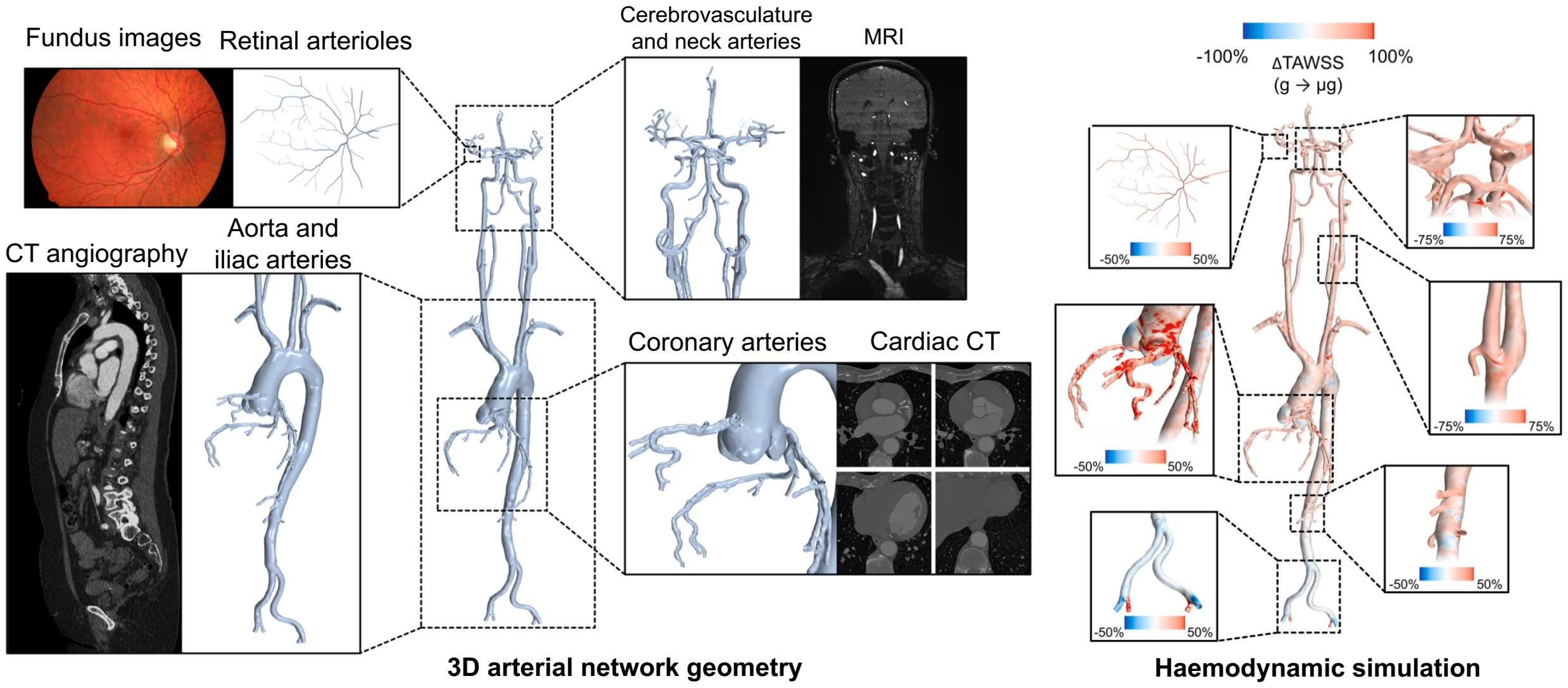

**Fig. 5**: Multi-organ digital twin. Here, we use the modeling of large scale artery hemodynamics from the heart to the eye as example. Image adopted from Caddy et al.[132] under Creative Commons license CC BY 4.0.

Beyond stroke and cardiac arrest, multi-organ digital twins are essential for modeling complex pathophysiological interactions in systemic diseases. The COVID-19 pandemic highlighted the profound and dynamic coupling between the respiratory and cardiovascular systems, where pulmonary inflammation and impaired gas exchange can precipitate secondary cardiac dysfunction. While statistical analyses of raw clinical data have failed to detect subtle COVID-19-related impairments in pulmonary circulation[147], integrated mathematical modeling has proven more insightful. For instance, calibrated 0D models of the cardiocirculatory system, when coupled with representations of pulmonary pathophysiology, have successfully quantified the hemodynamic consequences of COVID-19-related pneumonia, demonstrating the unique power of mechanistic digital twins to reveal hidden cross system hemodynamics[148,149].

The principle of cross-system integration extends to other organ pairs. The cardiovascular system critically regulates ocular perfusion, while intraocular pressure (IOP) and retinal blood flow are essential for ocular health. Although retinal vascular characteristics are established indicators of cardiovascular status[150,151] and AI-based tools now leverage retinal images for non-invasive cardiovascular risk assessment[152], digital twin models of the heart and eye have been developed in isolation. To bridge this gap, frameworks such as the closed-loop Eye2Heart model have been proposed, simulating coupled cardiovascular-ocular dynamics (Fig. 5) to explore scenarios like IOP variation or changes in cardiac contractility[153]. Looking ahead, the potential for even broader integration is significant. Preliminary multi-omics analyses reveal strong correlations in imaging-derived phenotypes, genetics, and proteomics among the brain, heart, and eye[154], suggesting that brain-heart-eye digital twins could offer unprecedented insights into their tripartite physiological coupling and shared disease mechanisms.

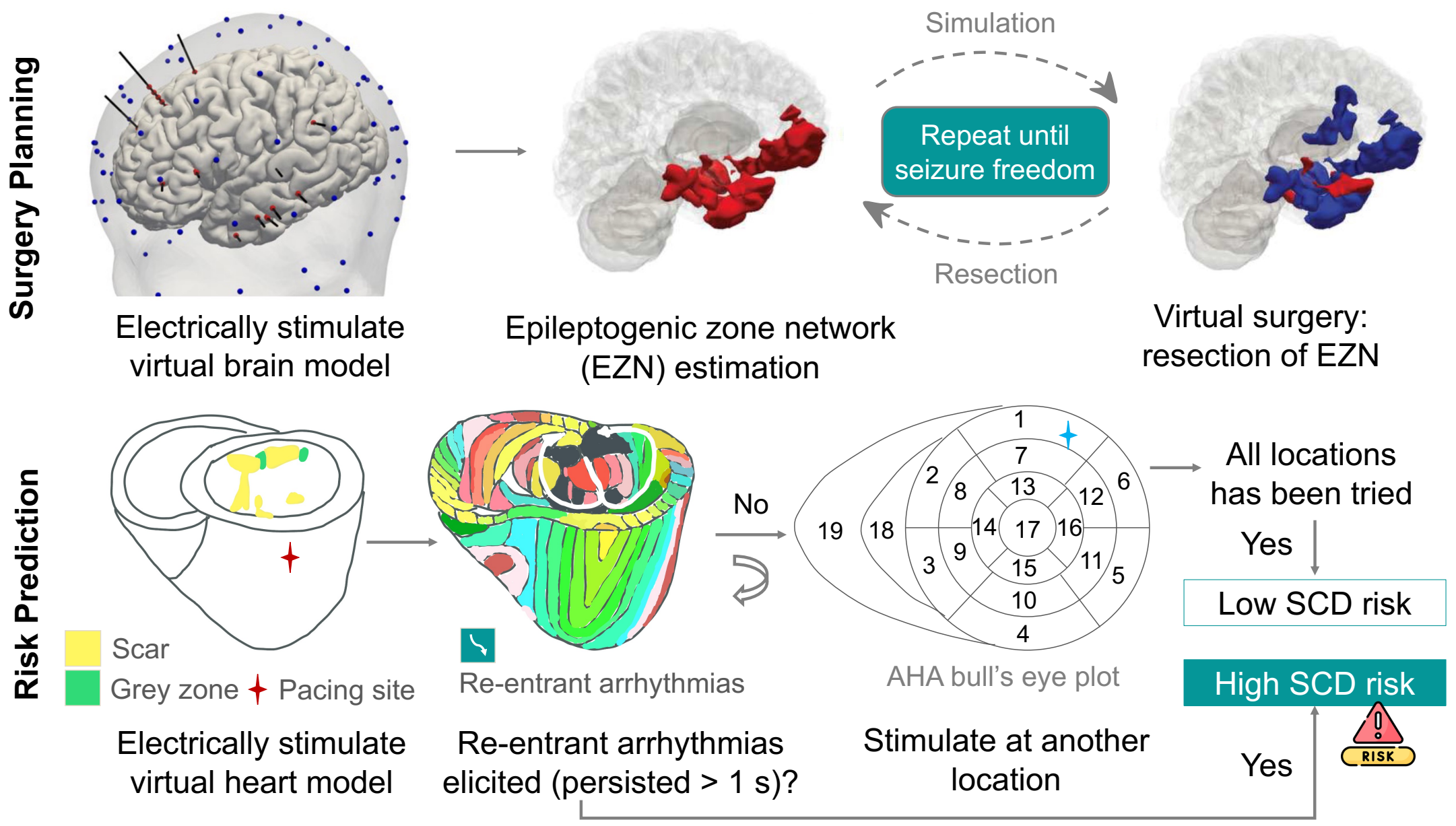

**Fig.6**: Illustration of clinical applications of digital twin for personalized healthcare. Here, we use the personalized modeling in drug-resistant epilepsy and personalized sudden cardiac death risk (SCD) prediction as examples. Image adopted from ref. 13 and ref. 133 under Creative Commons license CC BY 4.0.

## 4.3 Digital Twin-Guided Personalized Medicine

The ultimate validation of organ digital twin technology lies in its capacity to improve clinical decision-making and patient outcomes. While widespread adoption is still emergent, several pioneering applications demonstrate the translational potential of this paradigm (Fig. 6). These successes typically involve a closed-loop workflow: integration of multimodal patient data to personalize a biophysical model, execution of in-silico trials to predict treatment response, and translation of the virtual insights into a clinical action. The most mature examples are found in cardiology and neurology, where model personalization is supported by rich functional imaging and electrophysiological data. In cardiovascular medicine, cardiac digital twins have advanced from research to clinical decision support, particularly in complex arrhythmia management[133,155,156]. The most notable clinical application is the planning of catheter ablation for atrial fibrillation (AF) using patient-specific electrophysiological models. The optimal ablation strategy is determined through an iterative computational process: the digital twin simulates AF to locate initial ablation sites, perform virtual ablation, and then test the AF inducibility. This virtual ablation-and-test cycle repeats until AF can no longer be sustained, defining a final, personalized lesion set. In a proof-of-concept study with ten AF patients, this model-guided approach significantly improved procedural success rates and reduced operative time compared to standard strategies[156]. A second, rapidly advancing application is in transcatheter aortic valve implantation, where biomechanical twins of the aortic root derived from CT are used to simulate device deployment, predict paravalvular leak risk, and optimize valve sizing and positioning preoperatively[157].

In neurology, epilepsy surgery planning represents a landmark application of digital twinning. Personalized brain network models, or The Virtual Brain instances, are constructed from the structural and diffusion MRI of patients to simulate large-scale electrophysiological dynamics[158]. By incorporating the location of an epileptogenic lesion inferred from electroencephalography (EEG)/ magnetoencephalography (MEG)/ stereo-EEG, these models can simulate seizure propagation pathways. This allows neurosurgeons to virtually test the impact of different resection strategies on network dynamics, aiming to maximize seizure freedom while preserving critical cognitive function. Prospective clinical validation has shown that model-guided resection strategies can lead to improved surgical outcomes in drug-resistant epilepsy[13,159]. Beyond epilepsy, similar personalized brain modeling approaches are being explored to optimize deep brain stimulation parameters for Parkinson disease[160] and to predict functional recovery after stroke[161-163].

Beyond these established examples, digital twins are entering clinical pilot studies across other specialties, especially in terms of oncology[1]. For example, a triplenegative breast cancer digital twin has been developed to simulate the tumor growth and treatment response to neoadjuvant chemotherapy[164]. This digital twin framework, validated in a breast cancer cohort (n=105), achieved an AUC of 0.82 in predicting pathological complete response (pCR) and yielded a significant improvement in pCR rate of 20.95-24.76%. The common thread across these applications is a shift from a reactive, population-based medicine to a predictive, patient-specific paradigm, where therapeutic decisions are vetted in a risk-free virtual environment before being applied to the individual.

## 5 Outlook

This survey has outlined the state-of-the-art computational frameworks for building organ-specific digital twins, charting a path from foundational anatomical modeling to integrative functional simulation. While anatomical twinning has matured significantly, powered by medical foundation models and robust image analysis pipelines, the realization of fully functional digital twins remains a profound challenge. Current models are largely static, personalized in anatomy but not yet in dynamic function, due to the immense difficulty of acquiring longitudinal, time-series clinical data and the prohibitive computational cost of real-time, high-fidelity physics-based simulation. Here, AI, particularly in the form of hybrid physics-informed AI, emerges as a pivotal enabler. Such hybrid paradigms elegantly embed known biophysical laws while leveraging data-driven methods to infer unknown or patient-specific parameters, creating models that are both interpretable and adaptable. Presently, heart and brain digital twins are the most clinically advanced, demonstrating tangible impact in arrhythmia management and epilepsy surgery. However, the vision of a true, dynamically updating digital twin and the even more ambitious goal of an interconnected multi-organ system, remains too ambitious. Bridging this gap will require concerted advances in longitudinal data acquisition, real-time computational surrogates, and robust validation frameworks, ultimately steering the field from a promising technological concept toward a cornerstone of precision healthcare.